\documentclass[twocolumn,journal]{IEEEtran}
\usepackage{cite}
\usepackage{graphicx}
\usepackage{amsmath}
\usepackage{gettitlestring}
\begin{document}
\title{To Sense or Not To Sense}
\author{Ahmed El Shafie\\
\small \begin{tabular}{c}
Wireless Intelligent Networks Center (WINC), Nile University, Giza, Egypt. \\
\end{tabular}
}
\date{}
\maketitle
\begin{abstract}
A longer sensing time improves the sensing
performance; however, with a fixed frame size, the longer sensing time
will reduce the allowable data transmission time of the secondary
user (SU). In this paper, we try to address the tradeoff between sensing the primary channel for $\tau$ seconds of the time slot proceeded by randomly accessing it and randomly accessing primary channel without sensing to avoid wasting $\tau$ seconds in sensing.  The SU senses primary channel to exploit the periods of silence, if the primary user (PU) is declared to be idle the SU randomly accesses the channel with some access probability $a_s$. In addition to randomly accesses the channel if the PU is sensed to be idle, it possibly accesses it if the channel is declared to be busy with some access probability $b_s$. This is because the probability of false alarm and misdetection cause significant secondary throughput degradation and affect the PU QoS. We propose variable sensing duration schemes where the SU optimizes over the optimal sensing time to achieve the maximum stable throughput for both primary and secondary queues. The results reveal the performance gains of the proposed schemes over the conventional sensing scheme, i.e., the SU senses the primary channel for $\tau$ seconds and accesses with probability $1$ if the PU is declared to be idle. Also, the proposed schemes overcome random access without sensing scheme.

The theoretical and numerical results show that pairs of misdetection and false alarm probabilities may exist such that sensing the primary channel for very small duration overcomes sensing it for large portion of the time slot. In addition, for certain average arrival rate to the primary queue pairs of misdetection and false alarm probabilities may exist such that the random access without sensing overcomes the random access with long sensing duration.
\end{abstract}
\section{Introduction}
Secondary users (SUs) are seen as prime candidate for increasing spectrum efficiency. The SUs exploit periods of silence of primary users (PUs) under certain QoS for the PUs. In a typical cognitive radio setting the cognitive transmitter senses primary activity and decides on accessing the channel on the basis of the sensing outcome, which we refer to as conventional sensing scheme, $\mathcal{S}_{c}$. This approach is problematic because sensing may affect primary QoS. Spectrum sensing to detect the presence of the
PUs is, therefore, a fundamental requirement in cognitive
radio networks.

In a fixed frame size, the longer sensing time
will shorten the allowable data transmission time of the SU while improving the sensing
performance \cite{peh2009optimization}. Hence, a sensing-throughput tradeoff problem was formulated
in \cite{liang2008sensing} to find the optimal sensing time that maximizes the secondary
users' throughput while providing adequate protection to the
primary user. Both the sensing time and the
cooperative sensing scheme affect the spectrum sensing performance,
such as the probabilities of detection and false alarm. These probabilities
affect the throughput of the secondary users since they determine
the reusability of frequency bands \cite{peh2009optimization}. Recently, the authors of \cite{KarimSultan} proposed a random access scheme where the SU randomly accesses the primary channel with some access probability without employing any sensing scheme.

In this work, we try to address the impact of sensing the electromagnetic spectrum for $\tau$ seconds of the time slot proceeded by randomly accessing it with some access probability based on the sensing outcome. We investigate the maximum stable throughput of the users. We optimize over the sensing duration that the SU can use to maximize its maximum stable throughput. In addition, we try to address when the SU can switch between sensing the channel for $\tau$ seconds of the time slot and randomly accesses it without employing any sensing schemes.

The rest of the paper is organized as follows. Next we describe the system model adopted in this paper. The proposed scheme are discussed in Section \ref{sec3}. In Section \ref{sec5}, we provide some numerical results, and finally, we conclude the paper in Section \ref{sec6}.
\section{System Model}\label{sec2}
\subsection{MAC Layer}
\begin{figure}
  \includegraphics[width=1\columnwidth]{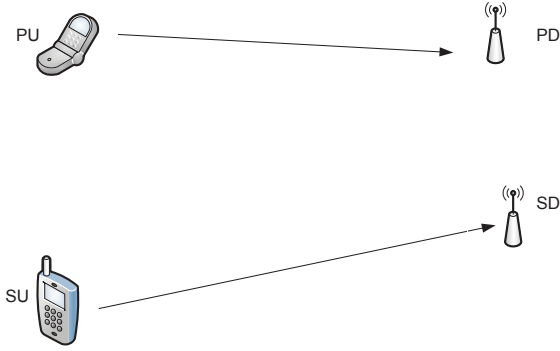}\\
  \caption{Primary and secondary links.}\label{fig1}
\end{figure}

Our network consists of one PU and one SU as depicted in Fig. \ref{fig1}. The SU senses the primary channel to detect the possible activities of the PU. If the PU is declared to be idle, the SU accesses the channel with some access probability $a_s$. The main contribution in this paper is that the SU randomly accesses the channel preceded by spectrum sensing for $\tau$ seconds of the time slot instead of accessing with probability one. First, we consider the case where the SU randomly accesses the channel only if the PU is declared to be idle. Secondly, we investigate the case where the SU randomly accesses the channel if the PU is sensed to be idle with access probability $a_s$ and if it is declared to be busy with probability $b_s$. This is because the SU tries to mitigate the impact of misdetection and false alarm probabilities.

The probability that the SU misdetects the primary activity is $P_{\rm MD}$ and the probability that the SU sensor generates false alarm is denoted as $P_{\rm FA}$. We assume that the primary transmitter has a buffer $Q_p$
to store the incoming traffic packets, while the secondary
transmitter has a buffer $Q_s$ to store its
own arrived traffic packets. We assume all buffers are of
infinite length. We consider time-slotted transmission where all packets have the same size and one time slot is sufficient for the transmission
of a single packet. The arrival processes of the primary and the secondary transmitters are assumed to be independent Bernoulli processes with mean arrival rates $\lambda_p$ and $\lambda_s$ packet per time slot, respectively.

\subsection{Physical Layer}
In this work, we characterize the success and failure of packet
reception by outage events and outage probability. The probability of outage event of the link between SU and secondary destination (SD), $P_{s,sd}$, can be calculated as in \cite{ShafieSultan}. The transmitter adjusts its transmission rate depending on when it starts transmission during the time slot.
Assuming that the number of bits in a packet is $b$ and the time slot duration is $T$, the transmission rate is
\begin{equation}
r=\frac{b}{T-\tau}.
\label{r_i}
\end{equation}
\noindent If transmission is preceded by a spectrum sensing period of $\tau$ units of time the time remaining for SU transmission is $T-\tau$. Consider the SU and its destination PD, an outage for the transmission occurs when the transmission rate exceeds channel capacity
\begin{equation}
P^{\left(\tau\right)}_{s,sd}={\rm Pr}\biggr \{r > W \log_{2}\left(1+\gamma_{s,sd} \alpha_{s,sd}\right)\biggr\}
\end{equation}
\noindent where $\tau\in[0,T]$, $W$ is the bandwidth of the channel, $\gamma_{s,sd}$ is the received SNR when the channel gain is equal to unity, and $\alpha_{s,sd}$ is the channel gain, which is exponentially distributed in the case of Rayleigh fading. The outage probability can be written as
\begin{equation}
P^{\left(\tau\right)}_{s,sd}={\rm Pr}\Big\{\alpha_{s,sd}<\frac{2^{\frac{r}{W}}-1}{\gamma_{s,sd}}\Big\}
\end{equation}
\noindent Assuming that the mean value of $\alpha_{s,sd}$ is $\overline{\alpha}_{s,sd}$,
\begin{equation}
P^{\left(\tau\right)}_{s,sd}=1-\exp\bigg(-\frac{2^{\frac{r}{W}}-1}{\gamma_{s,sd}\overline{\alpha}_{s,sd}}\bigg)
\end{equation}
\noindent Let $\overline{P}_{s,sd}=1-P_{s,sd}$\footnote{Throughout the paper $\overline{x}=1-x$.} be the probability of correct reception. It is therefore given by
\begin{equation}\label{correctreception}
\overline{P}^{\left(\tau\right)}_{s,sd}=\exp\bigg(-\frac{2^{\frac{b}{TW\left(1-\frac{\tau}{T}\right)}}-1}{\gamma_{s,sd}\overline{\alpha}_{s,sd}}\bigg)
\end{equation}
Note that the outage probability increases as $\tau$ increases. The probability of channel outage of the link between the PU and the primary destination (PD), $P_{p,pd}$, has a similar formula with $\tau=0$ and the associated parameters.
\begin{equation}\label{correctreception}
\overline{P}_{p,pd}=\exp\bigg(-\frac{2^{\frac{b}{TW}}-1}{\gamma_{p,pd}\overline{\alpha}_{p,pd}}\bigg)
\end{equation}
For simplicity of notations, we use $\overline{P}^{\left(\tau\right)}_{s,sd}=\overline{P}_{s,sd}$.
\subsection{Misdetection and False Alarm Probabilities}
In this paper, we adopt the formulas of misdetection and false alarm depicted in \cite{liang2008sensing}. For a target false alarm, $P_{\rm FA}$, the probability of misdetection is given by:
\begin{equation}
\begin{split}
P^{\left(\tau\right)}_{\rm MD}=1-Q\bigg(\frac{1}{\sqrt{2 \gamma+1}}\big(Q^{-1}(P_{\rm FA})-\sqrt{\tau f_s}\gamma\big)\bigg)
 \end{split}
\end{equation}
where $\gamma$ is the received SNR, $f_s$ the sampling frequency, and $Q(.)$ is the complementary distribution function of the standard Gaussian. Note that $P_{\rm MD}$ is a monotonically decreasing function of $\tau$. Again to simplify the notation we omit the superscript of $P^{\left(\tau\right)}_{\rm MD}$.
\subsection{Stability Analysis}
 Let us denote the queue sizes of the transmitting terminals at any time instant $t$ by $Q_i^t$. Then, $Q_i^t$ evolves according to
\begin{equation}\label{queue}
    Q_i^{t+1}=\bigr(Q_i^t-\mathcal{U}_i^t\bigr)^{+}+\mathcal{A}^t_i
\end{equation}
where $\mathcal{U}_i^t$ is the number of departures in time slot $t$. $\mathcal{A}^t_i$ denotes the number of arrivals in time slot $t$ and is a stationary process by assumption with finite mean $\mathcal{E}\{\mathcal{A}^t_i\}=\lambda_i$. The function $(.)^{+}$ is defined as $(x)^{+}=\max(x,0)$. We assume that
departures occur before arrivals, and the queue size is measured at the beginning of the time slot \cite{sadek}.

A fundamental performance measure of a communication network is the stability of its queues. We are interested in the queues size. More rigourously, stability can be defined as follows \cite{szpankowski1994stability,sadek}.

\emph{Definition:} Queue $i\in\{p,s\}$ is stable, if
\begin{equation}\label{stabilityeqn}
   \lim_{t \rightarrow \infty  }{\rm Pr}\{Q_i^t<y\}=F(y) \hbox{ and}  \lim_{y \rightarrow \infty} F(y)=1
\end{equation}
If the arrival and service processes are strictly stationary, then we can apply Loynes's theorem to check for stability conditions \cite{loynes1962stability,sadek}. This theorem states that if the arrival process and the service process of a queue are strictly stationary processes, and the average service rate is greater than the average arrival rate of the queue, then the queue is stable, otherwise the queue is unstable.

Due to queues interaction we consider the case of backlogged SU (the SU always has packets to send). Note that this system is a lower bound on the original system because we consider that the SU interferes with the PU each time slot.
\section{Proposed Schemes}\label{sec3}
\subsection{Conventional Spectrum Sensing $\mathcal{S}_c$}
In a conventional spectrum sensing scheme, the SU senses the channel for $\tau$ seconds from the beginning of the time slot to detect the possible activities of the PU. If the PU is sensed to be idle the SU transmit the packet at the head of its queue with probability one. If the channel is declared to be idle the SU doesn't transmit. The primary queue is served if the SU correctly detects its activity and the link between PU and PD is not in outage. While the secondary queue is served if the primary queue is empty, the channel between the SU and its respective receiver is not in outage. Note that the probability that the PU being empty is given by
\begin{equation}
\begin{split}
{\rm Pr}\bigg\{Q_p=0\bigg\}=1-\frac{\lambda_p}{\mu_p}
\end{split}
\end{equation}
Thus, the average service rates of the nodes in this system are given by:
\begin{equation}
\begin{split}
\mu_p&=\overline{P}_{p,pd}\overline{P}_{MD}
 \end{split}
\end{equation}
\begin{equation}
\begin{split}
\mu_s&=\overline{P}_{s,sd}\overline{P}_{FA} \bigg(1-\frac{\lambda_p}{\overline{P}_{p,pd}\overline{P}_{MD}}\bigg)
\end{split}
\end{equation}
The maximum stable throughput, for specific $\tau$, is given by:
\begin{equation}
\begin{split}
\mathcal{R}(\mathcal{S}_c|\tau)&\!=\!\biggr\{\!(\lambda_p,\lambda_s)\!:\!\lambda_s \!<\!\overline{P}_{s,sd}\overline{P}_{FA} \bigg(\!1\!-\!\frac{\lambda_p}{\overline{P}_{p,pd}\overline{P}_{MD}}\!\bigg)\!\biggr\}
\end{split}
\end{equation}
The stability region of a backlogged SU is given by:
\begin{equation}
\begin{split}
\mathcal{R}(\mathcal{S}_c)&=\bigcup_\tau \mathcal{R}(\mathcal{S}_c|\tau)
\end{split}
\end{equation}
According to  Loynes's theorem the condition on stability of the PU and the SU are given by:
\begin{equation}
\begin{split}
\lambda_i <\mu_i, \ \ \hbox{and $i\in \{p,s\}$}
\end{split}
\end{equation}
The union over all possible values of $\tau$, $\bigcup_\tau \mathcal{R}(\mathcal{S}_c|\tau)$, can be obtained by solving the following optimization problem:
\begin{equation}\label{100}
\begin{split}
& \max_{\tau} \,\,
\lambda_s=\overline{P}_{s,sd}\overline{P}_{FA} \bigg(1-\frac{\lambda_p}{\overline{P}_{p,pd}\overline{P}_{MD}}\bigg)\\
&\,\,{\rm s.t.} \,\,\,\,\  0 \le \frac{\tau}{T} \le 1,\ \\& \,\,\,\,\,\,\,\,\,\,\,\,\ \lambda_p\le\overline{P}_{p,pd}\overline{P}_{MD}
\end{split}
\end{equation}
\subsection{First Proposed Random Access Scheme $\mathcal{S}_1$}
In this subsection, we assume that the SU randomly accesses the channel if and only if the PU is declared to be idle. For the system with a backlogged SU, denoted as $\mathcal{S}_1$, a packet from the PU is served if the complement of the event that the SU detects primary transmission correctly and accesses the channel is true and the channel between PU and PD is not in outage. The average service rate of the PU is given by:
\begin{equation}
\begin{split}
\mu_p&=\overline{P}_{p,pd}\bigg(1-a_{s}P_{\rm MD}\bigg)
 \end{split}
\end{equation}
 Now consider the secondary queue. Given that the primary queue is empty, a packet from $Q_s$ is served if the SU detects the primary activity correctly, it decides to access the channel, and the channel between SU and SD is not in outage. Thus, the SU average service rate is given by:
\begin{equation}
\begin{split}
\mu_s&=a_s \overline{P}_{s,sd}\overline{P}_{FA} \bigg(1-\frac{\lambda_p}{\overline{P}_{p,pd}\bigg(1-a_{s}P_{\rm MD}\bigg)}\bigg)
\end{split}
\end{equation}
One method to characterize the closure of the rates pair $(\lambda_p,\lambda_s)$, to obtain the stability region, is to solve a constrained
optimization problem to find the maximum feasible $\lambda_s$ corresponding
to each feasible $\lambda_p$ as $a_s$ varies over $[0,1]$ and $\tau$ over $[0,1]$. For a fixed $\lambda_p$, the maximum
stable arrival rate for the secondary queue is given by solving the following optimization problem \cite{sadek}:
\begin{equation}
\begin{split}
& \max_{a_s,\tau} \,\,
\lambda_s=a_s \overline{P}_{s,sd}\overline{P}_{FA} \bigg(1-\frac{\lambda_p}{\overline{P}_{p,pd}\bigg(1-a_{s}P_{\rm MD}\bigg)}\bigg)\\
&\,\,{\rm s.t.} \,\,\,\,\  0 \le a_s,\frac{\tau}{T} \le 1,\ \ \lambda_p\le\overline{P}_{p,pd}\bigg(1-a_{s}P_{\rm MD}\bigg)
\end{split}
\end{equation}
For a fixed $\tau$, the optimization problem is \textbf{concave} and it can be readily solved using Lagrangian multipliers. The optimal access probability is given by:
\begin{equation}
\begin{split}
a^*_s=\max\biggr(\min\biggr(\frac{1-\sqrt{\frac{\lambda_p}{\overline{P}_{p,pd}}}}{P_{\rm MD}},1\biggr),0\biggr)
\end{split}
\end{equation}
The stability region of a backlogged SU for a fixed $\tau$ is given by:
\begin{equation}
\begin{split}
\mathcal{R}(\mathcal{S}_1|\tau)&=\biggr\{(\lambda_p,\lambda_s):\lambda_s <a^*_s \overline{P}_{s,sd}\\ & \,\,\,\,\,\,\,\,\,\,\,\,\,\,\,\,\ \overline{P}_{\rm FA} \bigg(1-\frac{\lambda_p}{\overline{P}_{p,pd}\bigg(1-a^*_{s}P_{\rm MD}\bigg)}\bigg)\biggr\}
\end{split}\label{e1}
\end{equation}
The stability region of a backlogged SU is given by:
\begin{equation}
\begin{split}
\mathcal{R}(\mathcal{S}_1)&=\bigcup_\tau \mathcal{R}(\mathcal{S}_1|\tau)
\end{split}
\end{equation}
\subsection{Second Proposed Random Access Scheme $\mathcal{S}_2$}
 In addition to the operation of the SU in the first proposed scheme, the SU randomly accesses the channel even if the PU is declared to be busy with some access probability $b_s$, this scheme is denoted as $\mathcal{S}_2$. This is useful to mitigate the impact of false alarm probability.  Given that the channel between PU and PD is not in outage, a packet from the primary queue $Q_p$ can be served in either one of the following events: 1) if the SU detects the primary activity correctly and decides not to access the channel (which happens with probability $\overline{a}_s$); or 2) if the SU misdetects the primary activity and decides not to access the channel (which happens with probability $\overline{b}_s$). The average service rate of the primary queue can be given by:
\begin{equation}
\begin{split}
\mu_p&\!=\! P_{\rm MD}\overline{a}_{s}\overline{P}_{p,pd}\!+\! \overline{P}_{\rm MD}\overline{b}_{s}\overline{P}_{p,pd}
\end{split}
\end{equation}
Given that the primary queue is empty, a packet from $Q_s$ is served in either one of the following events: 1) if the SU detects the primary activity correctly, it decides to access the channel, and the channel between SU and SD is not in outage; or 2) if SU's sensor generates false alarm, the SU decides to access the channel with probability $b_s$ and the channel between SU and SD is not in outage. The average service rate of the secondary queue can be given by:
\begin{equation}
\begin{split}
\mu_s&=\biggr[a_s \overline{P}_{s,sd}\overline{P}_{\rm FA}+ b_s \overline{P}_{s,sd}P_{\rm FA} \biggr] \bigg(1-\frac{\lambda_p}{\mu_p}\bigg)
\end{split}
\end{equation}
The maximum stable throughput is given by solving the following optimization problem:
\begin{equation}\label{200}
\begin{split}
& \max_{a_s,b_s,\tau} \,\,
\lambda_s=\biggr[a_s \overline{P}_{s,sd}\overline{P}_{\rm FA}+ b_s \overline{P}_{s,sd}P_{\rm FA} \biggr] \bigg(1-\frac{\lambda_p}{\mu_p}\bigg)\\
& \ \ \ {\rm s.t.} \  0 \! \le\! a_s,b_s,\frac{\tau}{T}\! \le \!1 \\& \,\,\,\,\,\,\,\,\,\,\,\,\,\,\,\ \lambda_p\! \le \! P_{\rm MD}\overline{a}_{s}\overline{P}_{p,pd}\!+\! \overline{P}_{\rm MD}\overline{b}_{s}\overline{P}_{p,pd}
\end{split}
\end{equation}
The problem can be reduced to
\begin{equation}
\begin{split}
&\max_{\mathcal{T},\tau} \ \ \mathcal{C}^\dagger \mathcal{T}+\lambda_p \frac{\mathcal{C}^\dagger \mathcal{T}}{\mathcal{D}^\dagger \mathcal{T}+\mathcal{F}}\\
&\,\,{\rm s.t.} \,\,\,\,\  0 \le a_s,b_s,\frac{\tau}{T} \le 1\ \\& \,\,\,\,\,\,\,\,\ \mathcal{D}^\dagger \mathcal{T}+\mathcal{F}\le 0
\label{form}
\end{split}
\end{equation}
 Where $\dagger$ denotes vector transposition, $\mathcal{F}\!=\! \lambda_p\!-\!\overline{P}_{p,pd}$, and
\begin{equation}
\begin{split}
  \mathcal{D}\! & =\!\biggr[\!\begin{array}{c}P_{\rm MD}  \overline{P}_{p,pd}\\ \overline{P}_{\rm MD} \overline{P}_{p,pd} \\ \end{array}\!\biggr] \ \\ \mathcal{T}\!&=\! \biggr[\!\begin{array}{c}
a_s\\
b_s\end{array}\!\biggr] \\ \ \mathcal{C}\!&=\!\biggr[\!\begin{array}{c}
\overline{P}_{s,sd}\overline{P}_{FA}\\
\overline{P}_{s,sd}P_{\rm FA}\end{array}\!\biggr]
 \end{split}
\end{equation}
Fixing $b_s$ and $\tau$, we have the following the optimization problem.
\begin{equation}
\begin{split}
& \max_{a_s} \,\,
a_s \overline{P}_{\rm FA}\!+\!\frac{\lambda_p}{\overline{P}_{p,pd}} \frac{a_s \overline{P}_{\rm FA}\!+\!b_s P_{\rm FA} }{a_sP_{\rm MD}\!-\! (\overline{P}_{\rm MD}\overline{b}_{s}\!+\!P_{\rm MD})}\\
&\,\,{\rm s.t.} \,\,\,\,\  0 \le a_s\le 1 \\ & \,\,\,\,\,\,\,\,\,\,\,\,\,\,\,\,\  a_s\le \frac{P_{\rm MD}+ \overline{P}_{\rm MD}\overline{b}_{s}-\frac{\lambda_p}{\overline{P}_{p,pd}}}{P_{\rm MD}}
\end{split}
\end{equation}
The optimization problem, given $b_s$ and $\tau$, is \textbf{concave} and can be readily solved. The solution is provided in the Appendix. From the Appendix, for fixed $b_s$ and $\tau$ the problem is feasible when $P_{\rm MD}\!+\! \overline{P}_{\rm MD}\overline{b}_{s}\!\ge\! \frac{\lambda_p}{\overline{P}_{p,pd}}$. The optimal value of $a_s$ is given in Eqn. (\ref{equation_long}).

\begin{figure*}[!t]
\normalsize
\setcounter{equation}{28}
\begin{equation}\label{equation_long}
\begin{split}
a_s^*&\!=\! \max\biggr\{\min\biggr\{\! \frac{\bigg(P_{\rm MD}+ \overline{P}_{\rm MD}\overline{b}_{s}\bigg)\!-\! \sqrt{\!\frac{\frac{\overline{P}_{\rm FA}\lambda_p}{\overline{P}_{p,pd}}\bigg(P_{\rm MD}+ \overline{P}_{\rm MD}\overline{b}_{s}\bigg)\!+\!P_{\rm MD}\frac{P_{\rm FA}\lambda_p}{\overline{P}_{p,pd}}}{\overline{P}_{\rm FA}}}}{P_{\rm MD}},\frac{P_{\rm MD}+ \overline{P}_{\rm MD}\overline{b}_{s}-\frac{\lambda_p}{\overline{P}_{p,pd}}}{P_{\rm MD}},1\biggr\},0\biggr\}
\end{split}
\end{equation}
\hrulefill
\vspace*{1pt}
\end{figure*}
The maximum stable throughput for a fixed $\tau$ is given by:
\begin{equation}
\begin{split}
\mathcal{R}(\mathcal{S}_2|\tau)&=\biggr\{(\lambda_p,\lambda_s):\lambda_s <\mathcal{C}^\dagger \mathcal{T}^*+\lambda_p \frac{\mathcal{C}^\dagger \mathcal{T}^*}{\mathcal{D} \mathcal{T}^*+\mathcal{F}}\biggr\}
\end{split}\label{eqn5}
\end{equation}
The maximum stable throughput of $\mathcal{S}_2$ is given by the union over all possible values of $\tau$
\begin{equation}
\begin{split}
\mathcal{R}(\mathcal{S}_2)&=\bigcup_\tau \mathcal{R}(\mathcal{S}_2|\tau)
\end{split}
\end{equation}
Note that $\mathcal{S}_1\bigcup \mathcal{S}_2=\mathcal{S}_2$ because $\mathcal{S}_1$ is achieved from $\mathcal{S}_2$ by setting $b_s=0$.
\subsection{Random Access without Sensing Scheme $\mathcal{S}_\circ$}
In this system, denoted as $\mathcal{S}_\circ$, the SU accesses primary channel without employing any sensing schemes. The average primary and secondary service rates are given by:
\begin{equation}
\begin{split}
\mu_p&= \overline{a}_{s}\overline{P}_{p,pd} \\ \mu_s&=a_s \overline{P}_{s,sd} \bigg(1-\frac{\lambda_p}{\mu_p}\bigg)
\end{split}
\end{equation}
As in \cite{KarimSultan} the maximum stable throughput of a backlogged SU, after including the channels outage probability, is given by:
\begin{equation}
\begin{split}
\mathcal{R}(\mathcal{S}_\circ)&=\biggr\{(\lambda_p,\lambda_s):\lambda_s < \overline{P}_{s,sd}\bigg(1-\sqrt{\frac{\lambda_p}{\overline{P}_{p,pd}}}\bigg)^2\biggr\}
\end{split}\label{e3}
\end{equation}
Note that in this scheme the probability of secondary channel outage is less than $\mathcal{S}_1$ and $\mathcal{S}_2$ because the SU doesn't waste $\tau$ seconds in sensing, i.e., $P_{s,sd}|_{\mathcal{S}_\circ}<P_{s,sd}|_{\mathcal{S}_j}$ and $j \in\{1,2\}$.

It should be noticed that given specific $\lambda_p$ and $\tau$, we possibly can find certain values of false alarm and misdetection probabilities such that when the SU randomly accesses the channel without sensing is better than randomly accessing preceded by sensing it for $\tau$ seconds and vice versa. The pair $(P_{\rm FA},P_{\rm MD})$ can be fully specified by the boundary points of the stability region of the schemes, i.e., using Eqns. (\ref{e1}), (\ref{eqn5}), and (\ref{e3}). In other words, the pair exists if and only if it satisfies the condition
 \begin{equation}
\begin{split}
\lambda_s|_{\tau,\lambda_p}\big(\mathcal{S}_2)<\lambda_s|_{\lambda_p}\big(\mathcal{S}_\circ\big)
\end{split}
\end{equation}
In addition, for a fixed $\tau_1$, one possibly can find $\tau_2>\tau_1$ for each $\lambda_p$, such that
 \begin{equation}
\begin{split}
\lambda_s|_{\tau_1,\lambda_p}\big(\mathcal{S}_2)>\lambda_s|_{\tau_2,\lambda_p}\big(\mathcal{S}_2)
\end{split}
\end{equation}
The union between the proposed schemes, $\mathcal{S}^{\left(o\right)}\!=\!\mathcal{S}_\circ \bigcup \mathcal{S}_1 \bigcup \mathcal{S}_2\!=\!\mathcal{S}_\circ \bigcup \mathcal{S}_2$, can be achieved by using a switching optimization parameter to switch from one scheme to another in order to maximize the maximum stable throughput of the SU given certain $\lambda_p$, $P_{\rm FA}$, $P_{\rm MD}$, and channels outage.
\section{Numerical results}\label{sec5}
\begin{figure}
\center
  \includegraphics[width=0.9\columnwidth]{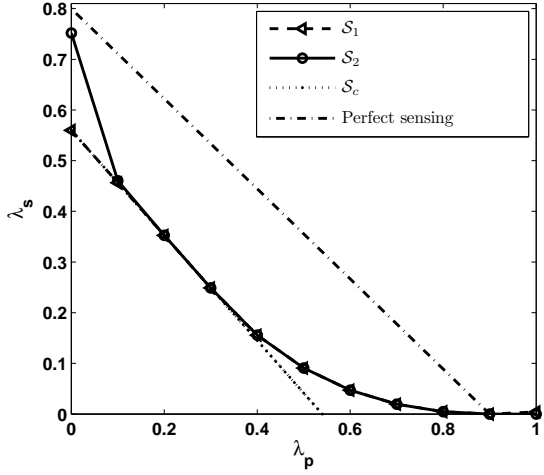}\\
 \caption{Stability region of the proposed system. The parameters used to generate the figure are: $P_{\rm MD}=0.3$, $P_{\rm FA}=0.2$, $\overline{P}_{p,pd}=0.9$, and $\overline{P}_{s,sd}=0.8$.}\label{fig8}
\end{figure}

\begin{figure}
\center
  \includegraphics[width=0.9\columnwidth]{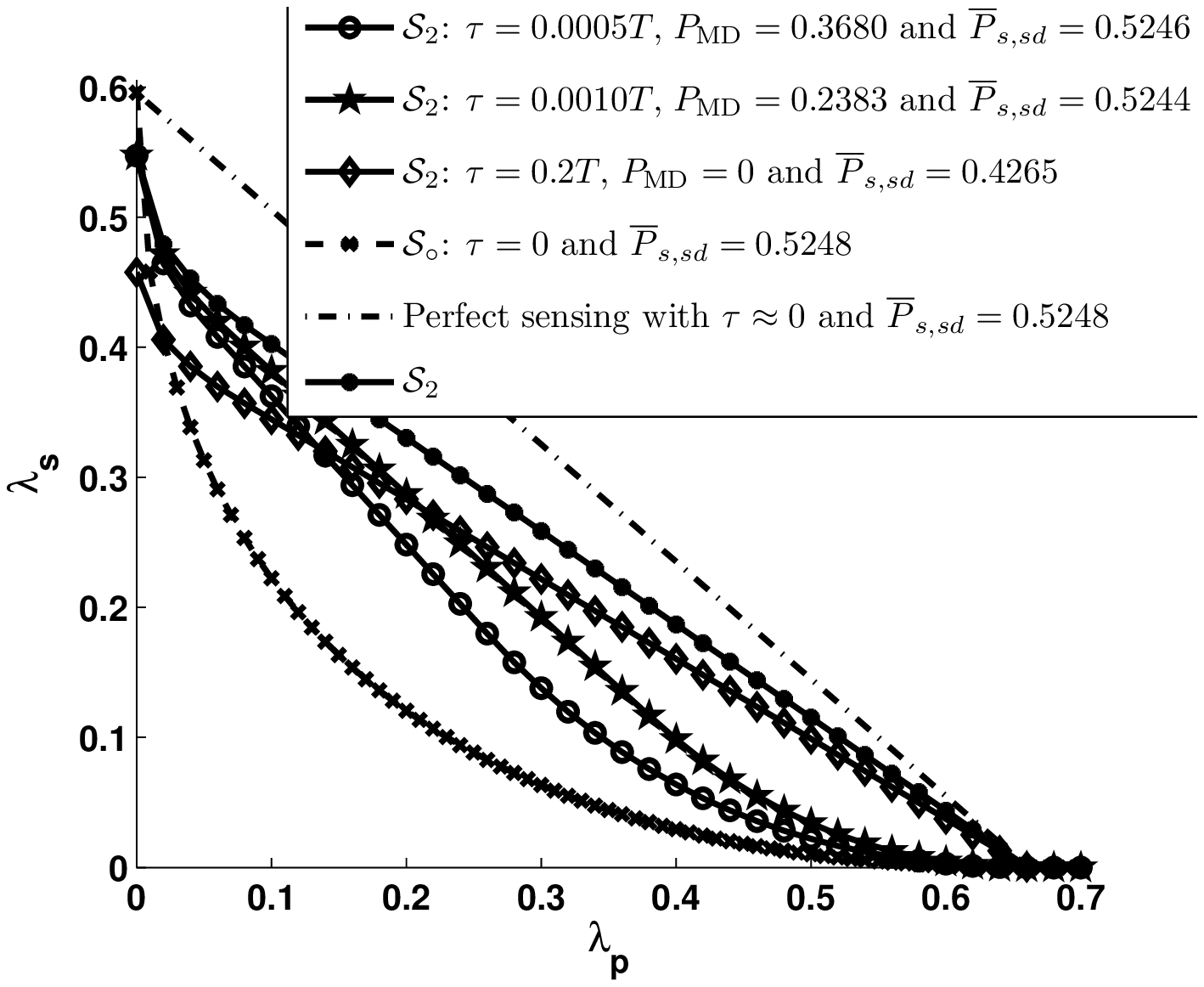}\\
  \caption{Stability region of the second proposed scheme, $\mathcal{S}_2$, as the sensing duration varies. The parameters used to generate the figure are: $P_{\rm FA}=0.2$ and $\overline{P}_{p,pd}=0.6609$.}\label{fig5}
\end{figure}

\begin{figure}
\center
  \includegraphics[width=0.9\columnwidth]{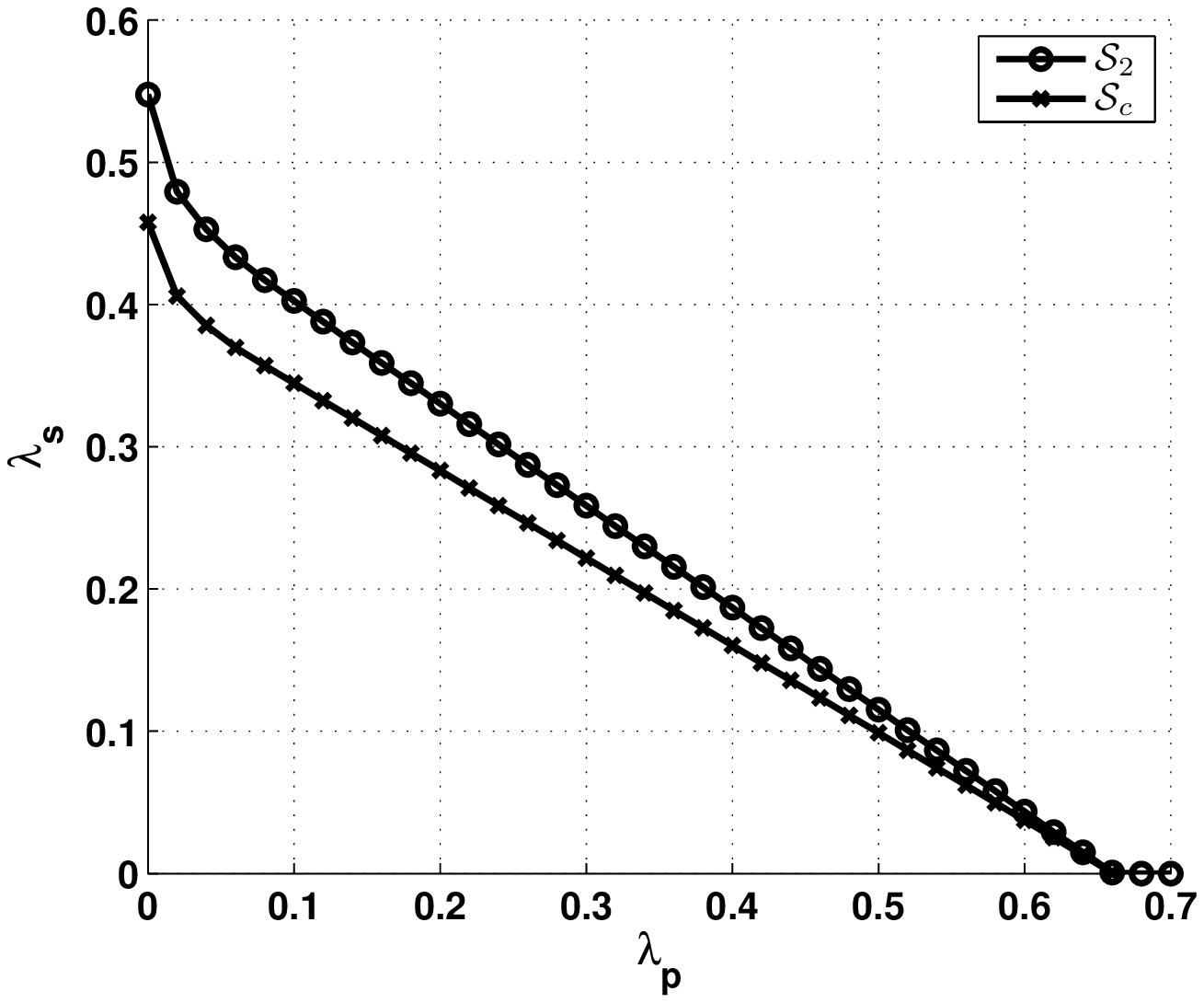}\\
  \caption{Stability region of the second proposed scheme, $\mathcal{S}_2$, as the sensing duration varies. The parameters used to generate the figure are: $P_{\rm FA}=0.2$ and $\overline{P}_{p,pd}=0.6609$.}\label{fig6}
\end{figure}

\begin{figure}
\center
  \includegraphics[width=0.9\columnwidth]{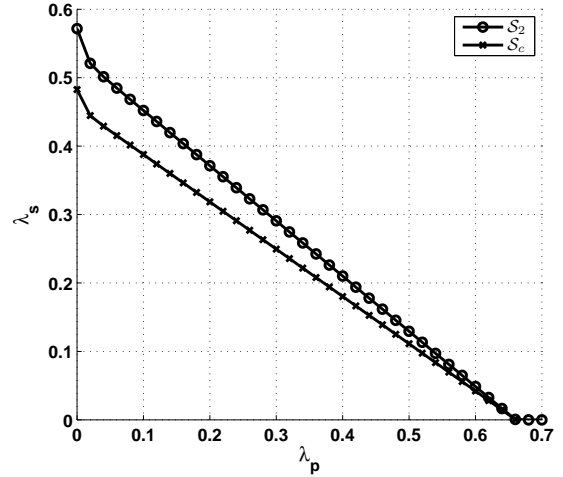}\\
  \caption{Stability region of the second proposed scheme, $\mathcal{S}_2$, as the sensing duration varies. The parameters used to generate the figure are: $P_{\rm FA}=0.1$ and $\overline{P}_{p,pd}=0.6609$.}\label{fig7}
\end{figure}

The maximum stable throughput of the considered schemes is shown in Figs. \ref{fig8} and \ref{fig5} reveal the expansion in the stability region of $\mathcal{S}_2$ as sensing duration varies. Fig \ref{fig5} shows the performance gain of $\mathcal{S}_2$ over $\mathcal{S}_c$ for different value of sensing duration. Also, the figure shows the stability region of $\mathcal{S}_2$ by taking the union over all possible sensing durations. Note that for small primary average arrival rate $\mathcal{S}_2$ with very small sensing duration overcomes $\mathcal{S}_\circ$ and $\mathcal{S}_2$ with long sensing duration. Also, $\mathcal{S}_\circ$ is better than $\mathcal{S}_2$ with long sensing duration because the SU doesn't waste $\tau$ seconds of transmission in sensing. Fig \ref{fig6} and \ref{fig7}  provide the solutions of the optimization problems (\ref{100}) and (\ref{200}) for different false alarm probability. It is a comparison between $\mathcal{S}_c$ and $\mathcal{S}_2$.

\section{Conclusion}\label{sec6}
In this work, we investigated the gains on the stability region of a SU randomly accesses the primary channel after making some sensing process. The results reveal the gains of the proposed schemes over the conventional sensing scheme and over the random access without sensing. The SU average service rate for the second proposed scheme with very small sensing duration can overcome sensing channel for long duration. We proposed variable sensing duration schemes where the SU optimizes over the optimal sensing time to achieve the maximum stable throughput for both primary and secondary queues. The results reveal the performance gains of the proposed schemes over the conventional sensing scheme. Also, the proposed schemes overcome random access without sensing scheme. The theoretical and numerical results show that pairs of misdetection and false alarm probabilities may exist such that sensing the primary channel for very small duration overcomes sensing it for large portion of the time slot. In addition, for certain average arrival rate to the primary queue pairs of misdetection and false alarm probabilities may exist such that the random access without sensing overcomes the random access with long sensing duration. For very low misdetection and false alarm probabilities the proposed schemes are reduced to the conventional scheme, i.e., if sensing outcome is robust all schemes coincide. A switching optimization parameter can be used to switch from one scheme to another based on the maximum stable throughput.

\section*{Appendix}
In this Appendix, we provide the solution of the following optimization problem:
\begin{equation}
\begin{split}
    & \max_{x}\ \ \ \frac{a\ x +f}{c\ x - d}+K\ x, \\& {\rm s.t.} \   \ \ 0  \le  x  \le  \frac{d-w}{c},\  x \le 1
    \end{split}
\end{equation}
where $a,f,c,d,K$, and $w$ are \textbf{positive} constants. For the problem to be feasible $d$ should be greater than or equal $w$, i.e., $d\ge w$. The first derivative of the objective function gives:
\begin{eqnarray}
    \frac{a(c\ x-d)-c \ (a\ x +f)}{(c\ x -d)^2}+K=0
\end{eqnarray}
After some mathematical manipulation
\begin{eqnarray}
(c\ x -d)^2=\frac{ad+cf}{K}
\end{eqnarray}
The roots of the quadratic equation are given by:
\begin{eqnarray}
x_1\!=\! \frac{d\!+\! \sqrt{\!\frac{ad\!+\!cf}{K}}}{c}, \ x_2\!=\! \frac{d\!-\! \sqrt{\!\frac{ad\!+\!cf}{K}}}{c}
\end{eqnarray}
One of the solutions is greater than the constraints which is $x_1>\frac{d-w}{c}$, thus $x^*=x_2$.
Including the constraints that $x\le 1$ and $x\ge 0$, thus the optimal value of $x$ is given by:
\begin{equation}
\begin{split}
x^*&\!=\! \max\biggr\{\min\biggr\{\! \frac{d\!-\! \sqrt{\!\frac{ad\!+\!cf}{K}}}{c},\frac{d\!-\!w}{c},1\biggr\},0\biggr\}
\end{split}
\end{equation}
with $d\ge w$.
\bibliographystyle{IEEEtran}
\bibliography{IEEEabrv,energy_bib}
\end{document}